\documentclass{article}

\usepackage{arxiv}

\usepackage[utf8]{inputenc} 
\usepackage[T1]{fontenc}    
\usepackage{hyperref}       
\usepackage{url}            
\usepackage{booktabs}       
\usepackage{amsfonts}       
\usepackage{nicefrac}       
\usepackage{microtype}      

\usepackage{authblk}

\usepackage{xcolor} \usepackage{ulem} \usepackage{changebar}

\usepackage{natbib}
\usepackage{graphicx} 
\usepackage{caption} 
\usepackage{subcaption} 

\usepackage{chngpage}
\usepackage{booktabs}
\usepackage{pdflscape}
\usepackage{afterpage}
\usepackage{capt-of}
\usepackage{adjustbox}
\usepackage{array}
\usepackage{amsmath}

\usepackage{textcomp}

\title{Identification of a Minimoon Fireball}

\author[1]{P.M. Shober \thanks{patrick.shober@postgrad.curtin.edu.au}}
\author[1]{T. Jansen-Sturgeon}
\author[1]{E.K. Sansom}
\author[1]{H.A.R. Devillepoix}
\author[1]{P.A. Bland}
\author[1]{M. Cup\'ak}
\author[1]{M.C. Towner}
\author[1]{R.M. Howie}
\author[1]{B.A.D. Hartig}
\affil[1]{Space Science and Technology Centre (SSTC), School of Earth and Planetary Sciences, Curtin University, GPO Box U1987, Perth, Western Australia 6845, Australia}

\begin{document}
\maketitle

\bibliographystyle{abbrvnat}

\begin{abstract}
Objects gravitationally captured by the Earth-Moon system are commonly called temporarily captured orbiters (TCOs), natural Earth satellites, or minimoons. TCOs are a crucially important subpopulation of near-Earth objects to understand because they are the easiest targets for future sample-return, redirection, or asteroid mining missions. Only one TCO has ever been observed telescopically, 2006 $RH_{120}$, and it orbited the Earth for about 11 months \citep{kwiatkowski2009photometry}. Additionally, only one TCO fireball has ever been observed prior to this study \citep{2016DPS....4831106C}. We present our observations of an extremely slow fireball (codename DN160822\_03) with an initial velocity of around $11.0\,\mbox{km s}^{-1}$ that was detected by 6 of the high-resolution digital fireball observatories located in the South Australian region of the Desert Fireball Network (DFN). Due to the inherent dynamics of the system, the probability of the meteoroid being temporarily captured before impact is extremely sensitive to its' initial velocity. We examine the sensitivity of the fireball's orbital history to the chosen triangulation method. We use the numerical integrator REBOUND to assess particle histories and assess the statistical origin of DN160822\_03. From our integrations we have found that the most probable capture time, velocity, semi-major axis, NEO group, and capture mechanism vary annually for this event. Most particles show that there is an increased capture probability during Earth's aphelion and perihelion. In the future, events like these may be detected ahead of time using telescopes like the LSST, and the pre-atmospheric trajectory can be verified.  
\end{abstract}

\keywords{asteroids, dynamics, meteorites, meteors}


\section{Introduction}

Occasionally when an object gets close to the Earth-Moon system, it is captured by the Earth's gravity. These objects are commonly called temporarily captured orbiters (TCOs), natural Earth satellites or “minimoons” \citep{2012Icar..218..262G}. The first mention of TCOs was by \citet{1913JRASC...7..145C} and then \citet{1916JRASC..10..294D} in a description of a extraordinarily long fireball that was witnessed over North America. Since the event lasted so long, according to witnesses, the source was speculated to be orbiting the Earth before entering the atmosphere. Besides this brief hypothesis, the study of TCOs was mostly left unexplored for the rest of the 20th century. During the space race, when artificial satellites began to be launched into orbit, there was speculation on whether or not natural Earth satellites would exist side-by-side with the artificial satellites \citep{1958Sci...128.1211B}.

For the last half-century, there have been many studies of captured objects by the large gas giants in the solar system, particularly Jupiter \citep{1977Icar...30..385H,1979Icar...37..587P,1996Icar..121..207K,2003AJ....126..398N,2007AJ....133.1962N}. There have also been several papers discussing the capture mechanisms and dynamics in the circular restricted three-body problem (CRTBP) and whether or not individual planets are even capable of sustaining a TCO population. Originally the models were simple and showed that only the large gas giants were capable of capturing satellites \citep{Yegorov1959}. Eventually, \citet{1972AJ.....77..177B} extended this methodology to any planet in the solar system. He showed that TCOs are possible for any planet when considering each in the limiting framework of the elliptic restricted three-body problem, instead of assuming circular orbits.

Following this study, \citet{1979CeMec..19..405C} was the first to explore the viability of a lunar assisted capture as a way to check for viable ballistic trajectories to the outer solar system’s planets. Since then, there have been a handful of studies interested in the feasibility of Moon-assisted captures along with using moons for decreasing the delta-V required for space missions to outer solar system objects \citep{2000Icar..148..139T,2002P&SS...50..269T,2011CeMDA.109...59L,2015Ap&SS.357..155G,2017CNSNS..48..211L}.

While studying the capture dynamics of Jupiter, several papers found that the capture duration was highly unpredictable \citep{1989AJ.....98.2346M,1996CeMDA..64...79B,1996Icar..121..207K}. This unpredictability was due to the fractal nature of the orbital phase space from which the objects originate. Furthermore, \citet{1989AJ.....98.2346M} stated that temporarily captured objects may have to have some chaotic origin, being on the boundary of two adjacent sinks (i.e., they can either evolve towards a heliocentric orbit or a planetocentric orbit). Thus, small perturbations in the initial conditions can radically change the evolutionary behavior of objects, i.e., whether or not it is captured and for how long the object is captured. \citet{2003Natur.423..264A} also showed that whether an orbit displayed prograde or retrograde behavior was intrinsically tied to the initial energy along with the size and distribution of regular satellites in the Hill sphere. This chaotic nature associated with the dynamics of natural satellites will make it much more difficult to predict where the meteoroid observed by the Desert Fireball Network (DFN) originated from in the solar system.

It was not until 2006 that the first Earth TCO was observed. Asteroid 2006 $RH_{120}$ orbited the Earth from July 2006 to July 2007 before escaping the Earth-moon system \citep{kwiatkowski2009photometry}. This asteroid is still the only observed TCO, but this will undoubtedly change once the Large Synoptic Survey Telescope (LSST) starts making regular survey observations in 2022 \citep{2008arXiv0805.2366I,2015IAUGA..2257052F}. \citet{2012Icar..218..262G} was the first to model TCOs that considered capture probability as a function of orbital element space for the NEO population. The model also calculated the size-frequency distribution and orbital distribution for TCOs. \citet{2017Icar..285...83F} expanded on this work by focusing on objects that approached the Earth and were captured but escaped before they could complete one orbit, also known as `temporarily captured flybys' (TCFs). Based upon these models, they predict that the largest object in orbit around the Earth at any given time is about 1 m in diameter and that these objects are typically captured through the Earth's co-linear  L1 and L2 regions. Additionally, they predicted that 0.1\% of all meteors were previously TCOs before they impacted the Earth. Given this information, we expect to find about 1 TCO within the DFN’s dataset.

\citet{2016DPS....4831106C} searched for fireballs that were natural satellites of the Earth before they impacted the atmosphere. They found one fireball detected by the European Fireball Network that had a 92-98\% chance of being captured by Earth before detection according to their model. Although, the capture duration for this meteoroid varied from 48 days up to over 5 years. \citet{2016DPS....4831106C} also looked at data from the Prairie Network in the US along with data collected by US Government sensors. None of the low-speed objects could be confidently said to be captured before impact due to the unknown or high uncertainty in the pre-atmospheric velocity for the measurements. To date, the event recorded by the EFN and described by \citet{2016DPS....4831106C} is the only fireball observed with a very high probability of originating from a TCO orbit. 

\citet{2012Icar..218..262G} assumed the orbit-density distribution is independent of the size-frequency distribution for their TCO model. While this is accurate for more substantial objects, it is unlikely true for smaller NEOs. The DFN and other fireball networks like it are particularly ideal for characterizing this portion of the meteoroid population. Using TCO fireball data collected from these types of networks, we can ascertain how likely the Granvik model is accurate for smaller size ranges. 

Generating an accurate orbital model for TCOs and TCFs is vital because these bodies are the most accessible in the solar system. They are the ideal targets for future sample-return, in-situ resource utilization (ISRU), and asteroid impact mitigation technology testing \citep{2014acm..conf...94C,2016P&SS..123....4B}. Additionally, since the average TCO orbits multiple times before escaping, this allows for multiple observations within a small time frame. These observations of TCOs can be used to understand the smallest members of the NEO population \citep{2014Icar..241..280B}. TCOs have the potential to have far-reaching effects on our understanding of asteroids and the history of the solar system along with many other future space-based technology applications. Thus, if we can better predict the orbital paths of these bodies based on observations and models, finding TCOs and TCFs will become easier.

The Desert Fireball Network (DFN) is a continental scale facility that observes fireballs in our atmosphere, calculates their pre-entry orbit, and determines where any possible meteorite material may land \citep{howie2017build}. There are currently 1300+ fully triangulated events detected by the DFN. Previous models of the natural Earth satellite population \citep{2012Icar..218..262G,2017Icar..285...83F}, predicted that about 0.1\% of all meteors impacting the Earth should have been temporarily captured prior to impact. Based on these models, assuming the orbit-density distribution is independent of the size-frequency distribution, there should be one or two events in the DFN dataset that were captured objects before impacting the atmosphere. 

\noindent
The questions to be addressed within this study include:
\begin{enumerate}
\item Is the number of TCOs in the DFN dataset consistent with previous models?
\item How would such meteoroids get captured by the Earth-Moon system and is this different than expected from past models?
\item How long might any TCOs have been captured before they hit the Earth?
\item How much does the presence of the Moon affect the capturability?
\end{enumerate}

\section{Event DN160822\_03 Observations} 
Within the orbital dataset of the DFN, one event was indeed flagged as a possible TCO: DN160822\_03. Here we will detail the event from initial observations to triangulation and will discuss in the following sections its' nature as a TCO. 

\paragraph{Event Detection} 

\begin{center}
\begin{tabular}{| c | c  c  c|}
\hline\hline
\textbf{observatory}  & range (km) $^*$ & start time (sec)$^\dagger$ & end time (sec)$^\dagger$ \\
\hline
Moolawatana           &       117       &            0.10            &           5.32           \\
Wertaloona            &       117       &            0.20            &           5.12           \\ 
Fowlers Gap           &       157       &            0.00            &           5.06           \\
Weekeroo              &       203       &            0.20            &           2.66           \\
Wilpoorinna           &       221       &            0.50            &           4.96           \\ 
Etadunna              &       270       &            1.10            &           4.16           \\ 
\hline
\end{tabular}
\captionof{table}{Locations and observation details for DFN observatories that detected event DN160822\_03. Start and end times are given relative to the event start/end (first event to detect fireball has relative start time of 0.00)\\
$^*$ Line of sight distance to start of trajectory\\
$^\dagger$ Relative to 12:17:10.826 UTC on 22 August 2016}
\end{center}

\begin{figure}
	\centering
	\includegraphics[height=12cm]{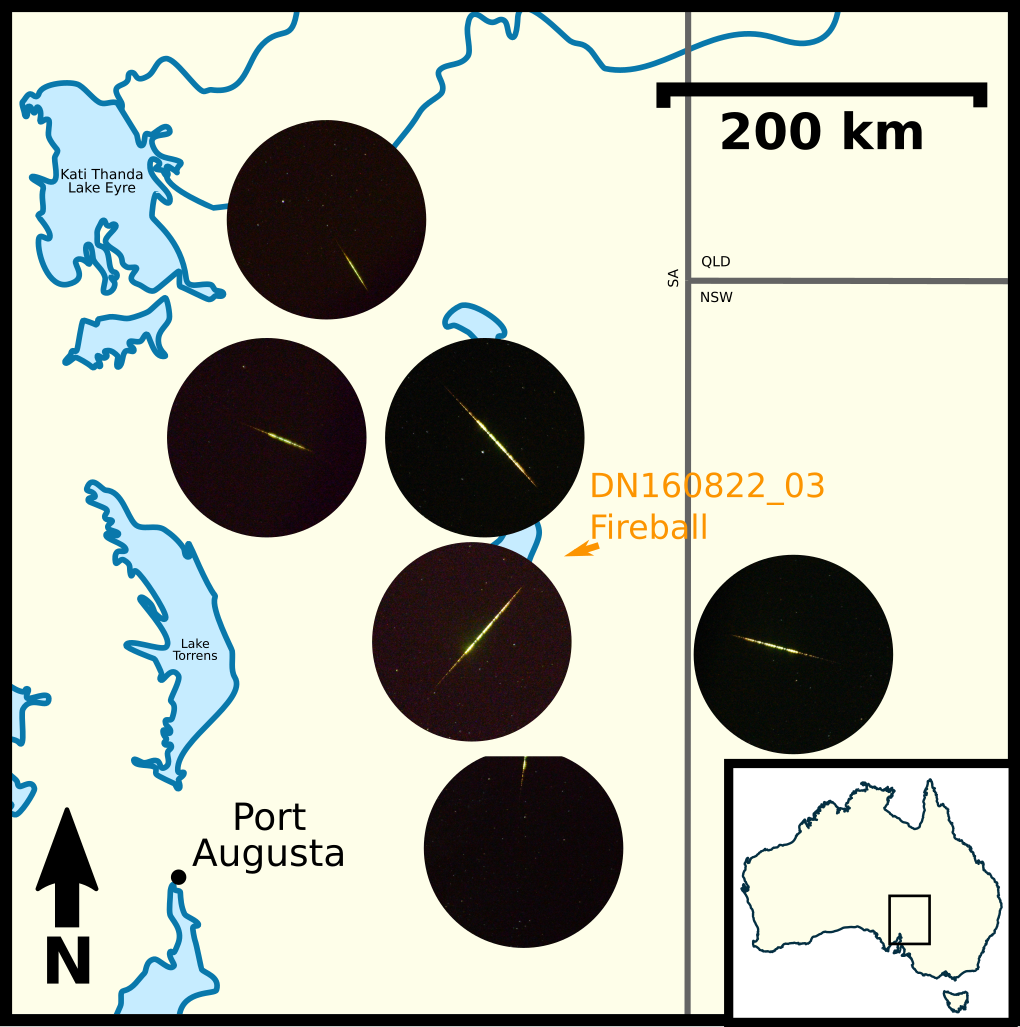}
	\caption{Map of camera observations for event DN160822\_03 in Southern Australia by the DFN. The orange arrow indicates the ground-track of the fireball's luminous trajectory. This path is extremely small due to the nearly vertical slope of the trajectory ($\approx 86.6^{\circ}$). Six camera observations were collected during the 5.32 sec duration.}
	\label{fig:map}
\end{figure}

DN160822\_03 was observed by six of the DFN's high-resolution fireball cameras in South Australia just before 11 PM local time on August 22, 2016 (Figure \ref{fig:map}). All but one of the cameras were able to image nearly the entire trajectory (Table 1). The event lasted over five seconds and had a nearly vertical atmospheric trajectory ($\sim87^{\circ}$). This high-angle impact argues against an artificial origin and pre-atmospheric trajectory integrations eliminate the possibility of standard satellite debris. Although, however unlikely, this does not eliminate the possibility of debris from Apollo or other past lunar/interplanetary missions. Table 2 summarizes the atmospheric trajectory, mass, and velocities determined for event DN160822\_03. 

The camera systems used to observe the event are described fully in \citet{howie2017build}. The absolute timing for the event was recorded using a de-Bruijn sequence that is encoded into the fireball image by using a liquid crystal shutter in addition to the built-in shutter \citep{howie2017submillisecond}. The liquid crystal is synchronized with a Global Navigation Satellite System (GNSS) module using a microcontroller, which produces absolute times accurate to $\pm0.4ms$. 

\begin{table}
\centering
\caption{Atmospheric trajectory of event DN160822\_03}
\begin{tabular}{| l  c  c |}
\hline\hline
                                        &      Beginning             &     Terminal              \\ 
\hline\hline 
Time (isot)                             &   2016-08-22T12:17:10.826  &  2016-08-22T12:17:16.146  \\
Height (km)                             &  74.1                      & 24.1                      \\ 
Mass (kg)                               &  $11.8$                    & $0.3$                     \\ 
Latitude (deg)                          &  -30.53009                 & -30.53960                 \\ 
Longitude (deg)                         &  140.38927                 & 140.36020                 \\ 
SLLS TOPS Velocity ($\mbox{km s}^{-1}$) &  $10.95\pm 0.07$           & $3.90\pm 0.18$            \\ 
DTF TOPS Velocity ($\mbox{km s}^{-1}$)  &  $11.07\pm 0.14$           & $3.77\pm 0.07$            \\
\hline\hline
RA (deg)                                & \multicolumn{2}{c|}{$-63.07\pm 0.00831$}               \\ 
Dec. (deg)                              & \multicolumn{2}{c|}{$-29.35\pm 0.00726$}               \\ 
Slope (deg)                             & \multicolumn{2}{c|}{$86.6\pm 0.01$}                    \\ 
Duration (sec)                          & \multicolumn{2}{c|}{5.32}                              \\ 
Best Convergence Angle (deg)            & \multicolumn{2}{c|}{87.8}                              \\ 
Number of Observations                  & \multicolumn{2}{c|}{6}                                 \\ 
Number of Datapoints                    & \multicolumn{2}{c|}{506}                               \\ 
\hline
\end{tabular}
\label{tab:obs}
\end{table}

\paragraph{Astrometric calibration} 

\begin{figure}
	\centering
	\includegraphics[height=10cm]{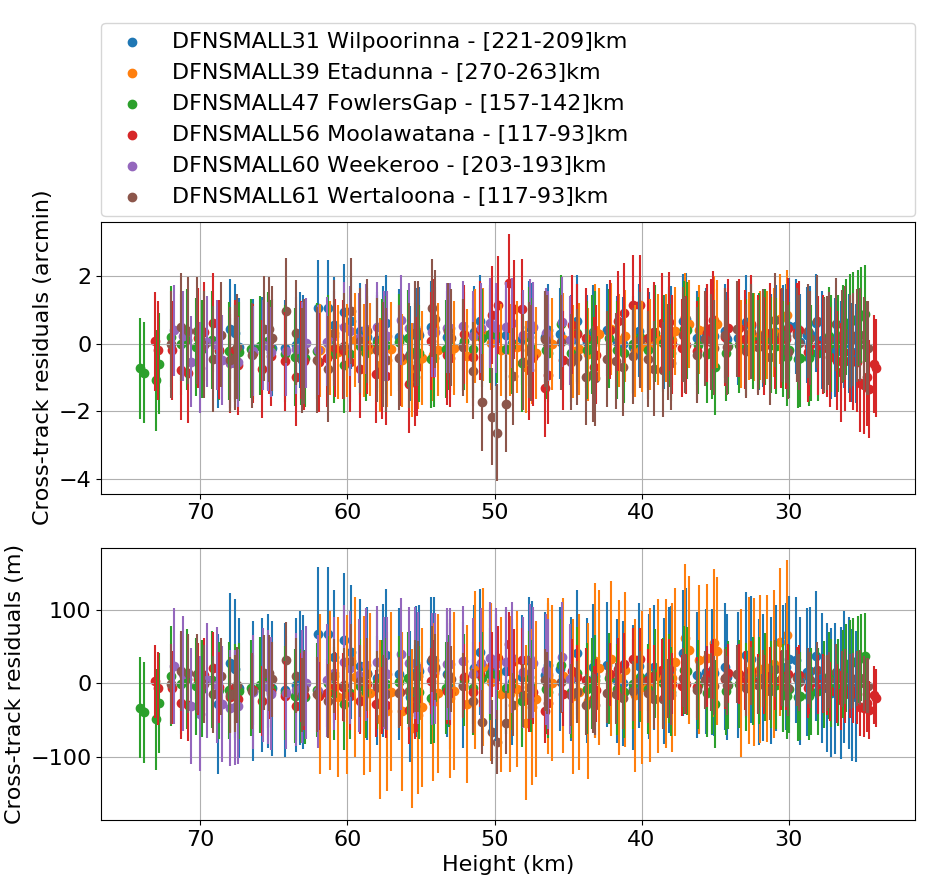}
	\caption{Cross-track residuals to the straight line trajectory fit (SLLS) of the event DN160822\_03. The dots correspond to the perpendicular distance between the observed lines-of-sight and the predicted straight line trajectory. The error bars represent the $1 \sigma$ formal astrometric uncertainties, however, these uncertainties are likely overestimated due to not well-constrained point-picking uncertainties (nominally $0.5$ pixel error). The observation range from each DFN station is given in the legend as [highest point - lowest point].}
	\label{fig:crosstrack_residuals}
\end{figure}

Astrometric calibration is performed using background stars, as described by \citet{2018M&PS...53.2212D}.
This results in astrometric measurements that are generally accurate ($1\sigma$) down to $\simeq 1.5$ minutes of arc (as shown by the errors-bars in Fig. \ref{fig:crosstrack_residuals}), limited by astrometric noise in this case.

\paragraph{Triangulation} 
During the analysis of the event detected by the DFN, two separate triangulation methods were used. We did this to check the sensitivity of the orbital history for this meteoroid to the triangulation method based on the work of previous studies \citep{10.1093/mnras/sty1841}. Our primary method is a straight line least squares (SLLS) algorithm, modified from \citep{borovicka1990comparison}, with an Extended Kalman Smoother (EKS) for velocity determination \citep{sansom2015novel}. Additionally, the Dynamic Trajectory Fit (DTF) of Jansen-Sturgeon, et al. (in prep.) was utilised alongside the traditional triangulation methods for comparison. The SLLS algorithm determines the straight-line trajectory by minimizing the angular distance between it and the observed lines-of-sight from every camera. The DTF algorithm is similar, however, it fits the observation rays to a trajectory based on meteor equations of motion, therefore dropping the straight-line assumption. One might say the SLLS is a purely geometric and simplifying fit, while the DTF is more based in reality. However, the initial velocity at the top of the luminous path ($v_{0}$) errors produced when using this DTF method cannot account for model error. The SLLS with an EKS velocity analysis can include this factor, therefore producing more reliable errors. Moreover, the event in question has a nearly vertical slope ($87.8^{\circ}$), and the luminous path deviates negligibly from linear (Fig. \ref{fig:crosstrack_residuals}). Thus, the backward integrations initiated after using the SLLS method in this paper are more statistically robust than those produced by the DTF method. We use both methods to demonstrate the highly sensitive pre-atmospheric orbit of event DN160822\_03 to the calculated $v_{0}$. 

\section{Methods}

\paragraph{Summary of Definitions and Abbreviations} 

Within this study we followed the notation of \citet{2012Icar..218..262G} and \citet{2017Icar..285...83F} for consistency (see Section \ref{sec:symbols} for a full list of symbols).
Consistent with \citet{2012Icar..218..262G} and  \citet{1996Icar..121..207K}, to be considered TC the particle has to be within at least 3 Hill radii of the Earth and have a planetocentric Keplerian energy $E < 0$. Additionally, to be classified as a TCO, the particle must have orbited the Earth at least once. Unlike previous studies \citep{2012Icar..218..262G,2017Icar..285...83F,2016DPS....4831106C}, instead of determining TCO membership by measuring the change in ecliptic longitudinal angle in the synodic frame, the TCO membership was determined by measuring the proportion of an orbital period each particle was captured. This reduces some of the ambiguity between TCFs and TCOs as demonstrated by \citet{urrutxua2017look}. 

\paragraph{Orbital Integrator} 
Simulations in this paper made use of the publicly available REBOUND code \footnote{http://github.com/hannorein/REBOUND}. REBOUND's 15th order IAS15 integrator was used for this study because of its resolution of close-encounters, its adaptive time-step, and the ability to incorporate non-gravitational forces along with other perturbations like the non-sphericity of the Earth \citep{2012A&A...537A.128R,2015MNRAS.446.1424R}. The IAS15 integrator is based on the RADAU-15 developed in \citet{1985dcto.proc..185E} used by \citet{2016DPS....4831106C} to model a captured-object impact detected by the European Fireball Network. IAS15 improves upon the RADAU-15 by suppressing the systematic error generated by the algorithm to well-below machine precision, implementing an adaptive time step, and adding the ability to include non-conservative forces easily while ensuring that the round-off errors are symmetric and at machine-precision \citep{2015MNRAS.446.1424R}.

\paragraph{Atmosphere Model} 
REBOUNDx's \footnote{https://github.com/dtamayo/REBOUNDx} publicly available additional forces were used as a way to add other forces to our model. We split up the regression model into two scripts: one that integrates back through the top of the atmosphere, and one that integrates back until the particles are out of the Earth-Moon system. The first integration code uses the whfast integrator provided by REBOUND along with the NRLMSISE-00 model 2001 \footnote{ported to python based off of Dominik Brodowski 20100516 version at http://www.brodo.de/english/pub/nrlmsis/} to take into account atmospheric drag that took place before the meteoroid started to ablate significantly in the upper atmosphere \citep{2015MNRAS.446.1424R}. The model produces a multivariate normal distribution of 10,000 particles given by our triangulation of the event. The particles vary in shape factor from a sphere to a brick (1.21-1.55) and are either chondritic or metallic in density ($3500\,\mbox{kg m}^{-3}$ or $7500\,\mbox{kg m}^{-3}$) \cite{gritsevich2009determination, consolmagno2008significance}. These particles are then integrated backward in time until all the particles are above 200 km. At this point, the simulation is handed-off to the next integration script. 

\paragraph{Integration Method} 
The long term integration script takes the distribution of particles from the results of the atmosphere script and generates particles from this distribution to be integrated out of the Earth-Moon system. The Sun, Moon, and Jupiter are directly added to the simulation from the JPL Horizons solar system data \footnote{https://ssd.jpl.nasa.gov/} and ephemeris computation service. Only these bodies were added to reduce the computational load and because they are the primary gravitational perturbers. REBOUNDx was used to incorporate orbital variations due to the Earth’s oblateness, and J2 and J4 gravitational harmonic coefficients were applied to the particles. We additionally accounted for radiation pressure using the REBOUNDx module. The model automatically adjusts the time-steps based on the non-linearity at that point in time. The integration itself is also split up into thousands of sections in order to save the appropriate outputs at regular time intervals. At the end of each integration section, the algorithm checks and records the particle's positions, orbital elements, and capture status, along with many other metrics. 

\begin{figure}
     \begin{subfigure}[b]{\textwidth}
         \centering
         \includegraphics[width=\textwidth]{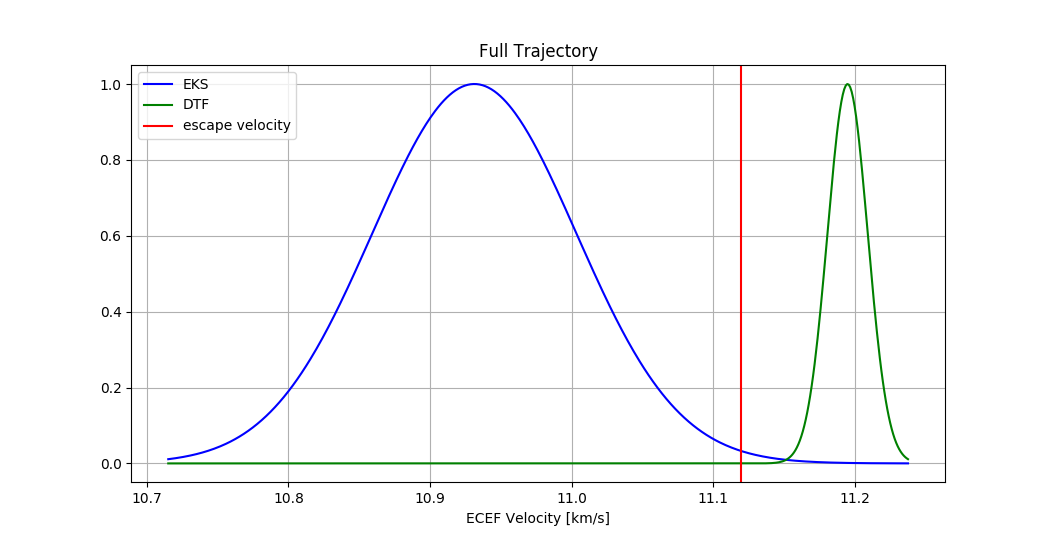}
         \caption{Full trajectory}
         \label{fig:sub1}
     \end{subfigure}
     \\
     \begin{subfigure}[b]{\textwidth}
         \centering
         \includegraphics[width=\textwidth]{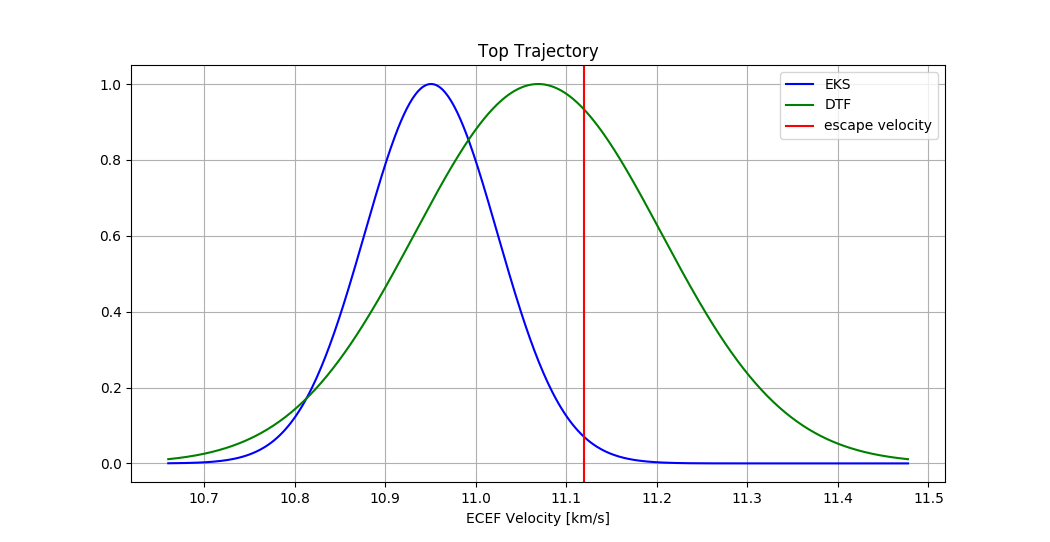}
         \caption{Top of trajectory}
         \label{fig:sub2}
     \end{subfigure}
        \caption{Comparison of the $v_{0}$ distribution generated by the EKS and the DTF methods using either (a) the full trajectory or (b) the top of the trajectory (observations $>$65 km altitude). Given the large amount of data collected for event DN160822\_03, 506 data points, the $v_{0}$ is more dependent than usual on the choice of triangulation and velocity determination methods. When only the top of the observed atmospheric trajectory is used, the models' assumptions affect the results less and the $v_{0}$ distributions converge.}
        \label{fig:distro graphs}
\end{figure}

In total, eight distinct orbit recursions were run. We varied the triangulation method, the meteoroid density, and the segment of the trajectory used to generate the orbits from the observations. In Table 3, we varied the density between `high' and `low', corresponding to metallic ($7500\,\mbox{kg m}^{-3}$) and chondritic ($3500\,\mbox{kg m}^{-3}$) densities respectively \citep{consolmagno2008significance}. Half of the orbital integrations were performed from triangulations using only the upper portion of the observed atmospheric trajectories. This `top of trajectory' (denoted `tops' in Table 3) is defined by all observations triangulated above 65 km altitude. This was done to reduce the dependency on the chosen triangulation model where high sample rates can observe variations due to additional physical effects occurring lower in the atmosphere (i.e., gravity, atmosphere). (Figure \ref{fig:distro graphs}). If a similar event occurred where the sampling rate was lower, varying the triangulation method could lead to an erroneous analysis of the results as the models will likely converge on full trajectory solutions.

By reducing the amount of data, the uncertainties increase and the mean TCO probabilities converge. Therefore, any study that states that a TCO fireball was observed based on atmospheric observations by photographic networks should be accepted with a degree of skepticism. Events like these, that come from inherently chaotic dynamics, cannot have their orbital histories definitively known. Usually, the triangulation and velocity determination methods do not vary the results significantly. Although, event DN160822\_03 is long-lasting, has a significantly large observational dataset (506 points, Table 2), and most importantly it is on the boundary of being geocentric and heliocentric. It is significantly more prone to model selection biases because slight variations in the starting conditions for this event drastically change the calculated orbital history. The particles were integrated back five years, enabling comparison with \citet{2016DPS....4831106C}.

\section{Results and Discussion}

\begin{figure}
]\centering
	\includegraphics[height=10cm]{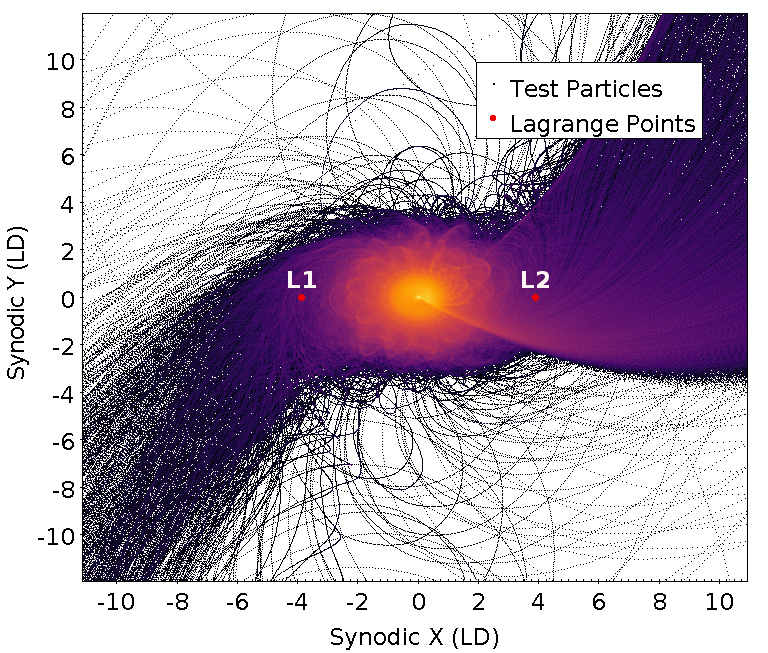}
	\caption{Particle orbits within the Sun-Earth-Particle synodic reference frame centered on the Earth’s center of mass and co-rotating with the Earth so that the direction of the sun is always at(-1AU, 0) in the x-y plane in this figure (not shown). The colors are indicative of the particles' spatial density, yellow being the most dense and black/purple being the least. The axes are in units of lunar distances (LD). There appears to be a clear preference of entry into the Earth-Moon system through either the L1 or L2 Lagrange points (represented by red points), as shown by the prevalence of trajectories in the directions of the co-linear Lagrange points.}
	\label{fig:synodic}
\end{figure}

\paragraph{Calculating Probabilities}
The capture probabilities listed in Table 3 were calculated in a very similar way to \citet{2016DPS....4831106C}. If a particle was deemed to be gravitationally captured while integrating backward, it was classified as a TCF until completing one orbit around the Earth and then it was reclassified as a TCO. The total number of TCs was determined by taking the sum of the TCO and TCF particles. If a particle appears to originate from the Earth (i.e., impacts the Earth in the backward integration), it would be removed from the TC, TCO, and TCF counts and classified as a `Sputnik'. Additionally, particles that were captured but never escaped from the Earth-Moon system within the 5 year integration time were labeled as still captured (SC). If the particles were never captured by the Earth-Moon, then they were marked as unbound (UB). Furthermore, if the particles passed within 3 or 1 lunar Hill spheres of the Moon, they were recorded as 3H or 1H respectively. 

Based on the pre-atmospheric orbit of event DN160822\_03, the probability that the meteoroid originated from typical artificial satellite debris is unlikely. However, due to the lack of spectral data, the possibility of originating from some previous lunar or interplanetary mission can not be eliminated. Subsequently, we have assumed based on the orbital characteristics that the event has a sufficiently small likelihood of coming from an artificial source. Thus, when calculating the capture probabilities, the Sputniks were removed from consideration due to their unlikelihood, producing the following general equation: 

\begin{subequations}
\begin{align}\label{eqn:eqlabel}
Probability_{min} &= \frac{Subset}{Total Particles - Sputniks}
\\
Probability_{max} &= \frac{Subset}{Total Particles - Sputniks - SC}
\end{align}
\end{subequations}

\noindent where the subsets are either SCs, TCs, TCOs, or TCFs. The SCs were considered invalid when calculating the TCF, TCO, and TC minimum percentages and included when calculating the maximum percentages. This is done because they could either eventually evolve into Sputniks or they could just have TCO dynamic lifetimes longer than the five year integration period. The 80,000 particles that describe this one event were integrated in groups of 1,000 for computational purposes, and the results of each run were very consistent with each other. The $\%SC$ was calculated using equation 1a.

\paragraph{Capture Probability}
Considering the large amount of data collected, the model choice affects the TC probability results more significantly when using the entire trajectory to determine $v_{0}$. In order to reduce this dependency of the model choice, the integrations were also performed using just the top of the observed atmospheric trajectory ($>$ 65 km altitude). This reduces the effect of the assumptions you make when choosing a model. Predictably, the two models' results tend to converge more when only the top is used (Fig. \ref{fig:distro graphs}).

During the integrations using the top of the trajectory, the particles generated from the SLLS still are nearly all either gravitationally captured or seem to originate from the Earth. On the other hand, about 30-60\% of the particles generated by the DTF method are TCs. The DTF produces non-conclusive probabilities for this event considering the $v_{0}$ distribution of the DTF is nearly centered (within $0.38\sigma$) on the escape velocity for the Earth at the corresponding altitude (Fig. \ref{fig:distro graphs}). In other words, the mean initial velocity (at the beginning of observations) predicted by the DTF method is very similar to the escape velocity. Therefore, the TCO probability for this event determined by the integrations initiated from DTF triangulation is predictably around $50\%$.

Given the results from the integrations (Table 3) using our most statistically robust triangulation method (SLLS with EKS), there is a $>$95\% probability that the meteoroid observed was captured by the Earth-Moon system before atmospheric entry (i.e., only $<$5\% chance it was heliocentric). Although, the pre-atmospheric path is impossible to exactly model due to the intrinsically chaotic nature of the system (as seen in Fig. \ref{fig:synodic}), and small variations in how the initial state of the fireball is determined has the potential to affect the resulting capture probability seriously. Especially considering that event DN160822\_03 probably had a close encounter with the Moon, producing chaotic scattering; the system is highly unpredictable. 

\begin{figure}
\centering
	\includegraphics[height=5cm]{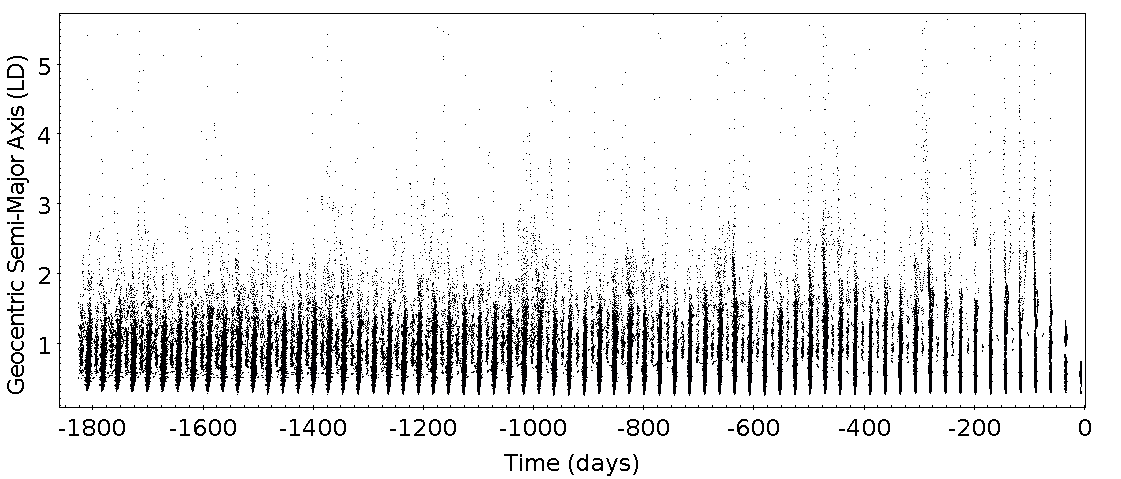}
	\caption{TCs that get within 3 Hill radii of the Moon (3H) produced by the SLLS and DTF. Each point represents one particle within 3H and the y-axis indicates the geocentric semi-major axis (LD) for that particle at that time. Most of the TCs in the simulations have close encounters with the Moon multiple times. The probability of an encounter increases once a month, due to the geometry of this specific event. This indicates that the Moon was likely critically important for the geocentric orbital evolution of the meteoroid and the impact of the meteoroid with the Earth.}
	\label{fig:3H_geo_a}
\end{figure}

\paragraph{Capture Mechanisms}
As exhibited in Table 3, the captured particles have a significantly higher amount of close encounters with the Moon compared to unbound particles. This implies that the Moon likely played a significant role in the meteoroid's eventual impact with the Earth. Considering nearly all of the particles generated from the SLLS/EKS are still captured at the end of the integration, this may imply that the meteoroid was an extremely long-lived TCO like those described in \citet{2012Icar..218..262G}. \citet{2012Icar..218..262G} found that the longest-lived TCO particles in their simulations were those that had multiple close encounters with the Moon, which lowered the apogee of the orbit below 1 LD. As seen in Fig. \ref{fig:3H_geo_a}, the temporarily captured particles within our simulations for the most extended times do indeed have numerous close encounters with the Moon throughout the integration. The presence of the Moon more often contributes to the length of the capture rather than the actual capture itself.

\begin{figure}
\centering
     \begin{subfigure}[c]{0.5\textwidth}
         \centering
         \includegraphics[width=\textwidth]{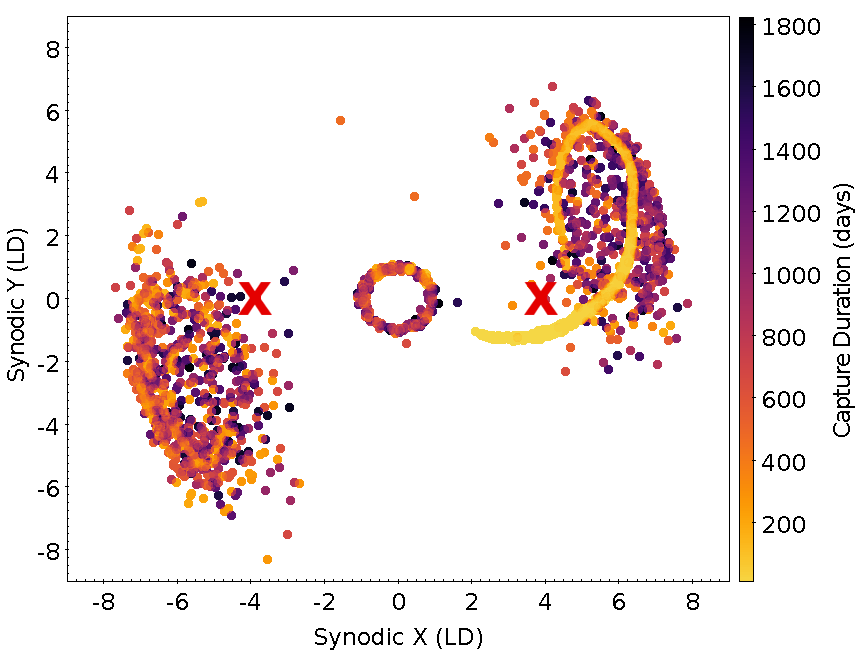}
         \caption{Variation in capture duration.}
         \label{fig:sub1}
     \end{subfigure}
     \\
     \begin{subfigure}[c]{0.5\textwidth}
         \centering
         \includegraphics[width=\textwidth]{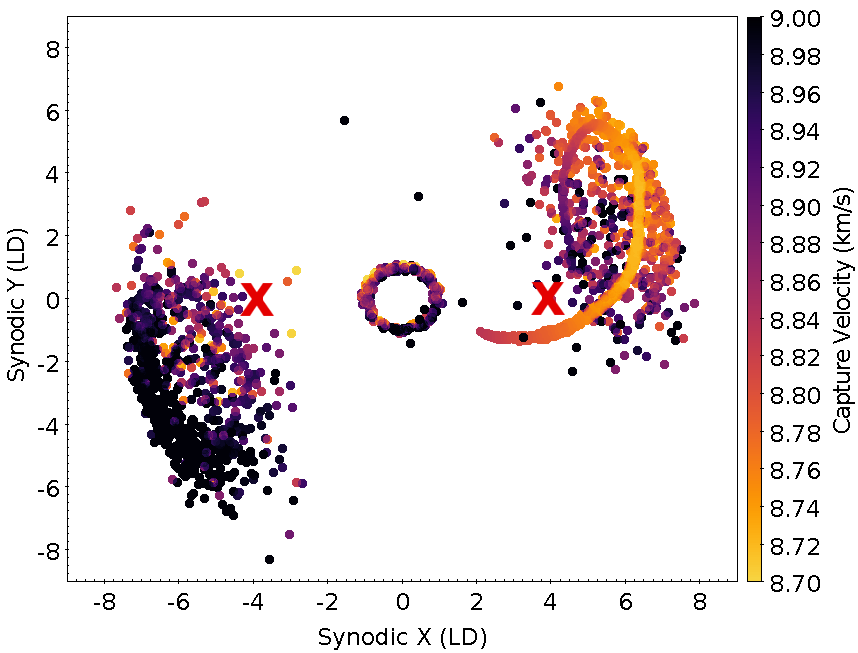}
         \caption{Variation in capture velocity.}
         \label{fig:sub2}
     \end{subfigure}
     \\
     \begin{subfigure}[c]{0.5\textwidth}
         \centering
         \includegraphics[width=\textwidth]{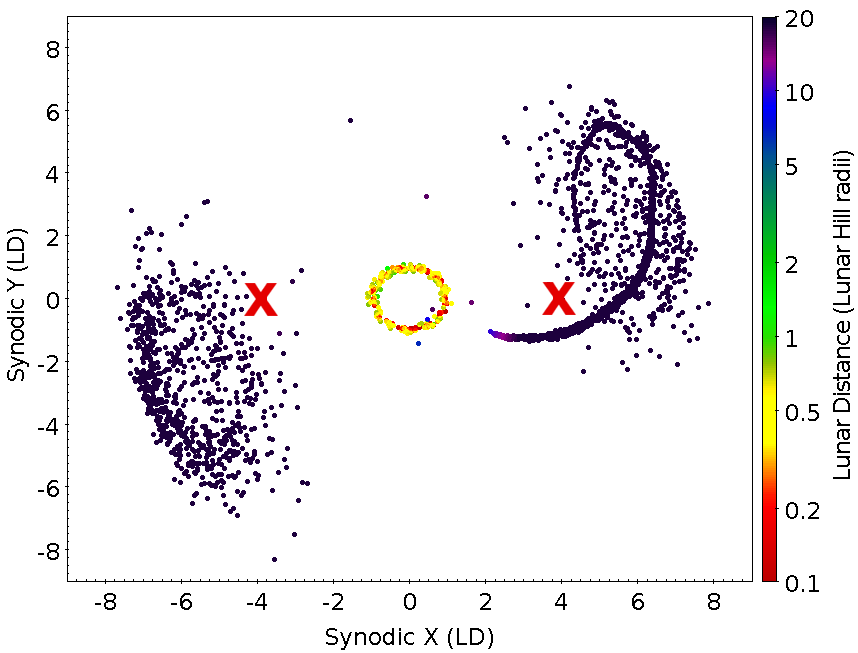}
         \caption{Variation in lunar distance.}
         \label{fig:sub3}
     \end{subfigure}
        \caption{Gravitational capture locations in synodic reference frame with L1 and L2 points marked by red crosses. The Sun-Earth synodic frame is centered on the Earth's center of mass and co-rotates with the Earth so that the direction of the sun in this case is always at (-1AU, 0) in the x-y plane. The figures above show 3 distinct capture regions: L1 capture, L2 capture and close lunar-encounter capture. The tail-like feature near the L2 point is caused by a large group of particles that were captured fairly quickly into the integration so they did not scatter as much.}
        \label{fig:capture graphs}
\end{figure}

In Fig. \ref{fig:capture graphs}, the capture distribution is clearly multi-modal. Most of the TCs are captured through the first or second Lagrange points, with the remaining TCs captured through a close encounter with the Moon. The capture location probabilities for the L1, L2, and lunar captures are $23.8\%$, $67.1\%$, and $9.1\%$ respectively. The specific Lagrange point capture locations depend on the Jacobi value for that given particle; in other words, the spread of Lagrange capture locations is due to the variations in the orbital energy of the particles. These capture mechanisms are easily seen in Fig. \ref{fig:capture graphs}. The capture locations also do not significantly change when the triangulation method is changed, however the proportion of the captures at each location does because of differences in the $v_{0}$ estimate in each model. 

\begin{figure}
     \begin{subfigure}[b]{\textwidth}
         \centering
         \includegraphics[width=\textwidth]{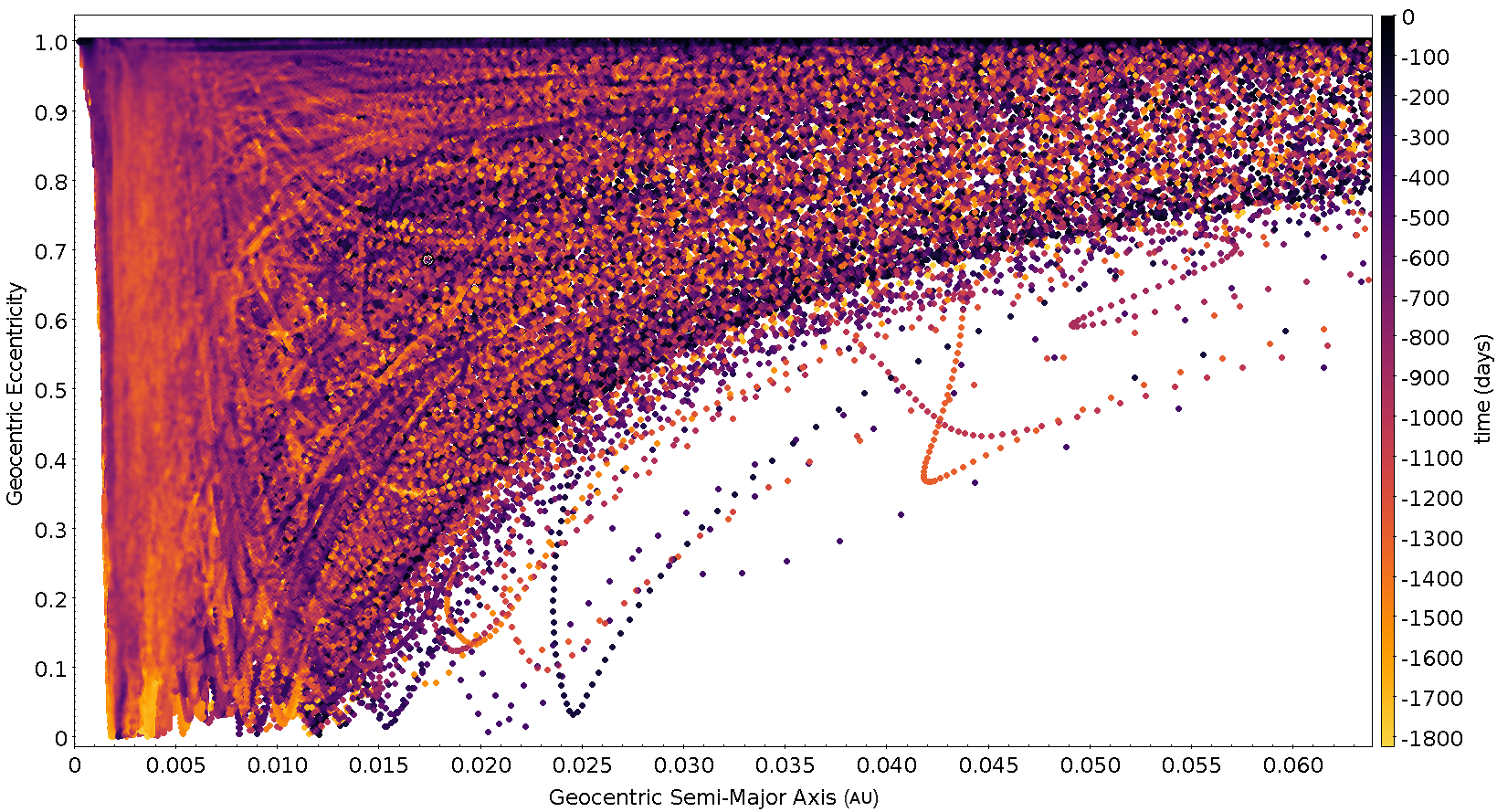}
         \caption{Semi-major axis vs Eccentricity}
         \label{fig:sub1}
     \end{subfigure}
     \\
     \begin{subfigure}[b]{\textwidth}
         \centering
         \includegraphics[width=\textwidth]{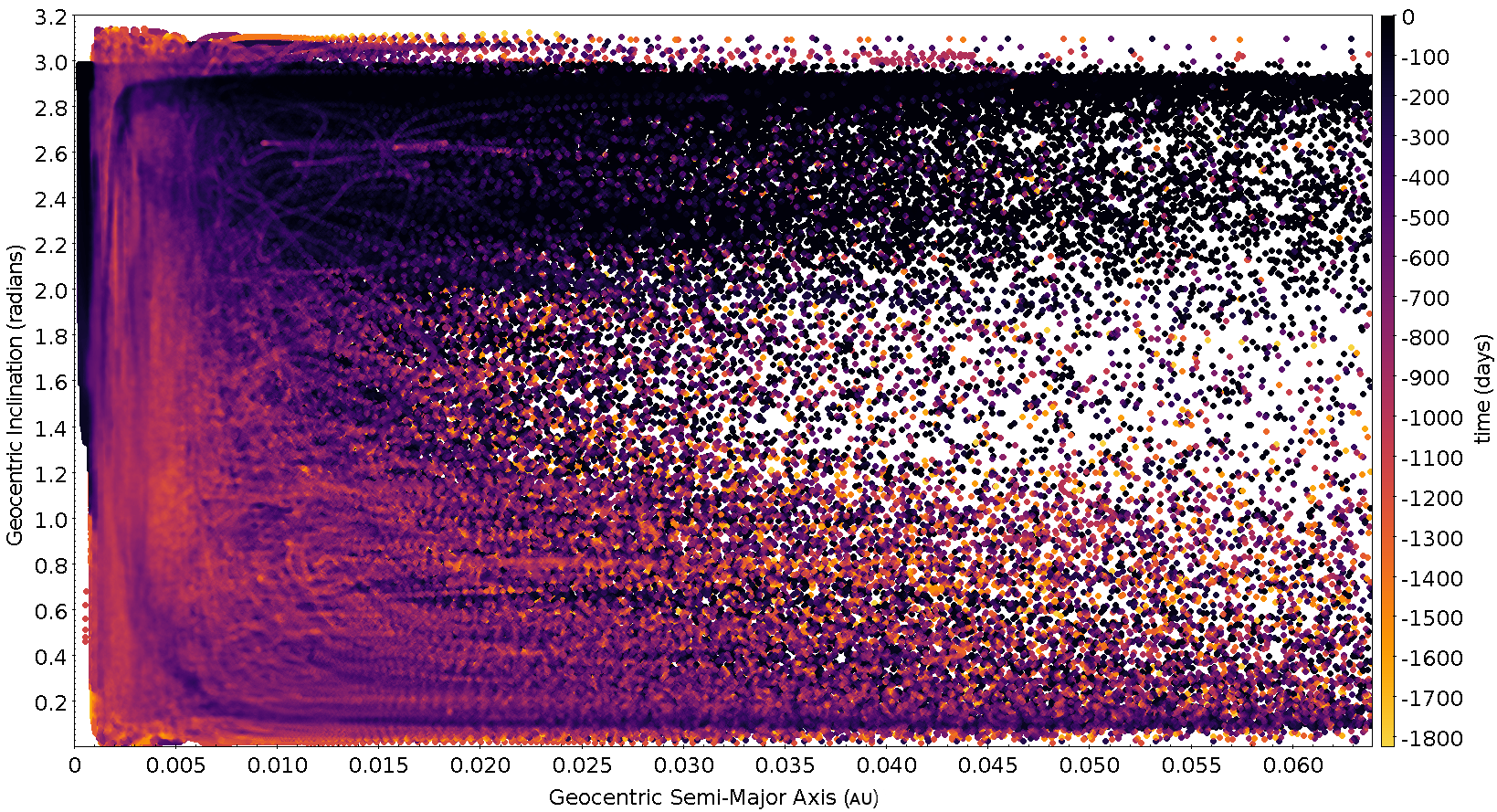}
         \caption{Semi-major axis vs Inclination}
         \label{fig:sub2}
     \end{subfigure}
        \caption{Geocentric semi-major axis vs eccentricity and inclination for the temporarily captured particles. The color bar is indicative of gravitational capture duration during the simulation. Yellow corresponding to a longer capture duration and black corresponding to a shorter capture duration. Particles that were captured the longest tended to have lower eccentricity, lower semi-major axis and lower inclinations. This is generally true because the particles that had more close encounters with the Moon tended to be less dynamically stable. Also, particles that were not able to transition from the initially highly eccentric detected orbit to lower eccentricities typically had lower capture durations. }
        \label{fig:goe graphs}
\end{figure}

\paragraph{Orbital Evolution}
As shown in Fig. \ref{fig:goe graphs}, there appear to be some trends over time for the geocentric orbital elements of captured particles. In Fig. \ref{fig:goe graphs}a, the captured particles that are integrated until they become heliocentric tend to approach higher semi-major axis and eccentricity values asymptotically. In Fig. \ref{fig:goe graphs}b, TCs that are retrograde and do not have a low semi-major axis encounter the Moon more often, causing them to be less dynamically stable and have shorter capture durations. The longest-lived particles have an apogee value lower than 1 LD, thus reducing the number of close encounters with the Moon. This is consistent with the longest-lived TCOs in the simulations done by \citet{2012Icar..218..262G} in which particles with the longest dynamical lifetimes tended to have multiple close encounters with the Moon that resulted in an orbit completely interior to the lunar orbit. Within this study, as shown in Fig. \ref{fig:goe graphs}, TCOs with low apogee values that had capture durations shorter than the integration period tended to evolve from highly eccentric retrograde orbits with larger semi-major axis values. This evolution from a retrograde, eccentric orbit to an orbit internal to the Moon was most likely due to a series of fortunate lunar close encounters like those described in \citet{2012Icar..218..262G}.

\clearpage
\begin{landscape}
\begin{center}

\begin{tabular}{| l | c | c | c | c | c | c | c |}
\hline\hline
 Triang. Method & Density & \# Sputniks & \%SC   & \% TCO      & \% TCF     & \% TC 3LH  & \% TC        \\ 
\hline\hline
SLLS full       & low     & 9728        & $98.1$ & $93.3-99.9$ & $0.1-6.2$  & $92.9$     & $99.5-100.0$ \\
SLLS full       & high    & 9711        & $96.4$ & $97.0-99.9$ & $0.1-2.9$  & $96.6$     & $100.0$      \\ 
\hline
SLLS tops       & low     & 9060        & $95.3$ & $88.5-99.5$ & $0.4-8.5$  & $87.8$     & $97.0-99.9$  \\ 
SLLS tops       & high    & 9173        & $95.5$ & $90.6-99.6$ & $0.3-6.8$  & $90.0$     & $97.4-99.9$  \\ 
DTF tops         & low     & 2974        & $36.5$ & $23.3-51.3$ & $8.4-13.3$ & $22.5$     & $36.6-59.7$  \\ 
DTF tops         & high    & 2879        & $35.6$ & $22.7-50.2$ & $8.5-13.2$ & $21.9$     & $35.9-58.7$  \\ 
\hline
\end{tabular}
\end{center}
\captionof{table}{Summary of 5 year recursion results for event DN160822\_03 in which over 16,000 valid particles were integrated, 10,000 for each run and 80,000 in total. TCs represent any captured particles, TCOs are captured and have orbited the Earth at least once, TCFs are captured and have not yet completed 1 orbit of the Earth, Sputniks are particles that originate from the Earth, SC represents particles that are still captured after 5 years, and TC 3LH is for TCs that go within 3 lunar Hill radii. The \%TCO, \%TCF, and \%TC values are calculated after removing Sputniks. In all of the integrations initialized from the SLLS, the Sputniks account for $>90\%$ of the particles. Due to the highly irregular orbit originating from the Earth, Sputnik particles are assumed to be invalid. There are no unbound particles that go within 3 lunar Hill radii recorded in the simulations, suggesting that the capture was facilitated by a close encounter with the Moon.}
\end{landscape}

\paragraph{Pre-capture Orbit}
By studying the trajectories of the simulated particles before encountering the Earth-Moon system, we find the event DN160822\_03 most likely to belong to the Apollo NEO group. Event DN160822\_03 produced particles that were 88.4\% Apollos, 6.2\% Amors, 2.9\% Atiras, and 2.5\% Atens. Although, due to the chaotic nature of the event, the heliocentric orbit is impossible to determine accurately without more data, preferably pre-atmospheric observations \citep{1989AJ.....98.2346M,2003Natur.423..264A}. 

\pagestyle{empty}
\pagestyle{plain}

\begin{figure}
     \begin{subfigure}[c]{\textwidth}
         \centering
         \includegraphics[width=0.9\textwidth]{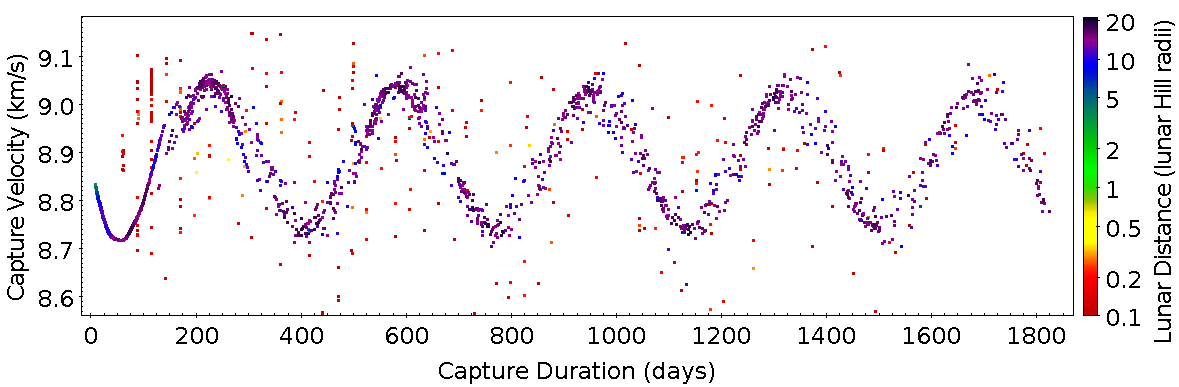}
         \caption{Variation in lunar distance.}
         \label{fig:sub1}
     \end{subfigure}
     \\
     \begin{subfigure}[c]{\textwidth}
         \centering
         \includegraphics[width=0.9\textwidth]{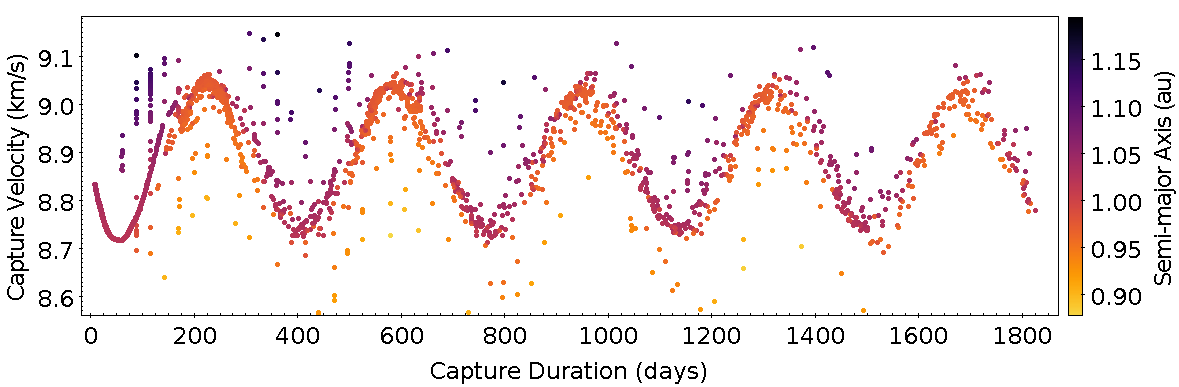}
         \caption{Variation in heliocentric semi-major axis.}
         \label{fig:sub2}
     \end{subfigure}
     \\
     \begin{subfigure}[c]{\textwidth}
         \centering
         \includegraphics[width=0.9\textwidth]{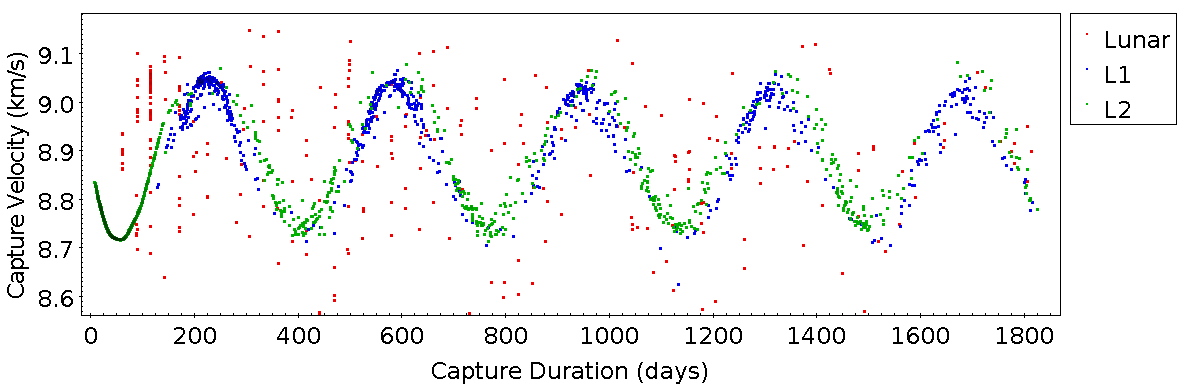}
         \caption{Variation in capture mechanism.}
         \label{fig:sub3}
     \end{subfigure}
     \\
     \begin{subfigure}[c]{\textwidth}
         \centering
         \includegraphics[width=0.9\textwidth]{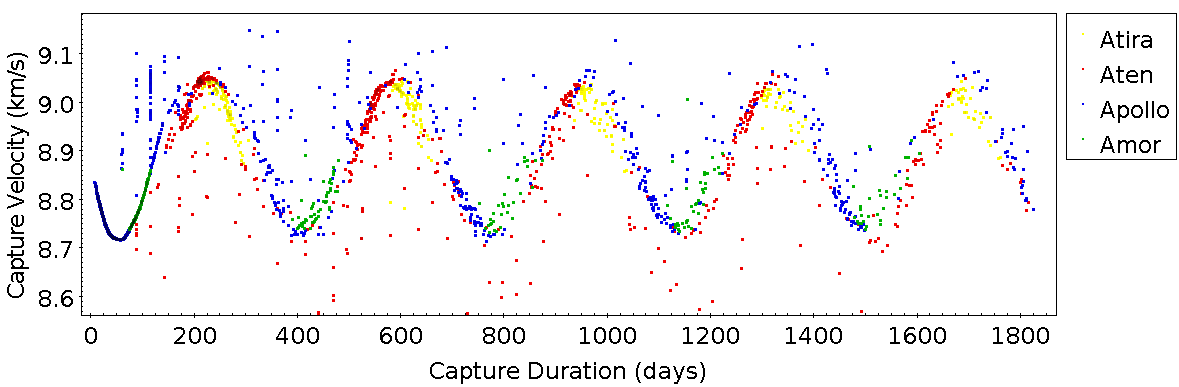}
         \caption{Variation in NEO type.}
         \label{fig:sub3}
     \end{subfigure}
        \caption{Total capture duration vs velocity during capture for TC particles. The relatively large annual variation in probable capture velocity results from the eccentricity of the Earth, as the Earth moves closer or further from the Sun during the year the capture velocity also varies. This annual variation in the probable capture velocity thus produces annual variations in the Lagrange point capture location and the source NEO group (Fig. b and Fig. c). Probably due to the geometry of the event (high eccentricity, low inclination, apogee $\approx1LD$), there also exists vertical bands of close lunar-encounter captures that occur every lunar month (Fig. a)}
        \label{fig:capvel graphs}
\end{figure}

\paragraph{Comparison to Models}
Finding this single TCO in the DFN dataset is consistent with the model of \citep{2012Icar..218..262G,2017Icar..285...83F}, although not statistically robust as numbers are small. We found the most probable capture locations were concentrated at the Earth's aphelion and perihelion, as described in both \citet{2012Icar..218..262G} and \citet{2017Icar..285...83F}. Although, as shown in Fig. \ref{fig:capvel graphs}, the particles captured in proximity to the L1 and L2 points clearly display an annual variation in the probable magnitude of the capture velocity. Also, unlike general models of the entire TCO population, particles were captured through close encounters with the Moon (Fig. \ref{fig:synodic}a) and had only slightly lower capture durations compared to the Lagrange point captures. 

Moreover, these close lunar encounter captures varied according to the lunar month for this event. This is seen in Fig. \ref{fig:capvel graphs}a, where the yellow/red points representing captures close to the Moon seem to make vertical stripes every ~28 days. This lunar cycle is also seen in Fig. \ref{fig:3H_geo_a}, where the amount of lunar encounters seems to spike every month. Although, this cycle of close lunar encounters every lunar month is most likely specific to the geometry of this event. Due to the low geocentric inclination and very high geocentric eccentricity, the particles generated are consistently capable of making numerous close encounters with the Moon. The presence of a lunar influence was also identified by \citet{2016DPS....4831106C}, where the lunar encounters tended to occur directly before the impact with the Earth. Implying the Moon is highly influential on whether or not TCOs dynamically evolve into an Earth-impacting orbit.

\paragraph{Annual Variations}
There is a relatively large annual variation in the expected capture velocity and capture semi-major axis, varying over 300 $ms^{-1}$ and 0.15 AU respectively for this particular event. This large annual variation in this event is due to the fact that the Earth does not have a perfectly circular orbit around the Sun. This eccentricity causes the L1 and L2 Lagrange points in the simplified circular restricted three-body problem to ``wobble" in and out throughout the year by about $3.4\%$ . As a result, the capture characteristics also ``wobble" throughout the year. This implies that the source region for TCs also varies annually with Atiras and Atens more likely to be gravitationally captured during perihelion (January) and Amors and Apollos more likely to be gravitationally captured during aphelion (July) (Fig. \ref{fig:capvel graphs}d). Atira and Aten orbits are more likely to be gravitationally captured during perihelion because the L1 and L2 points are closer to the Earth and faster objects relative to the Earth are capable of being captured, i.e., objects with orbits interior to that of the Earth. Conversely, the Amor and Apollos are more likely during aphelion because they orbit relatively more slowly and have orbits more outward from the Earth. As shown in Fig. \ref{fig:capvel graphs}d, interestingly the faster and slower lunar captures consistently come from Apollo and Aten type orbits respectively.  Additionally, this annual variation in probable capture velocity also implies that the capture mechanism by L1 and L2 varies annually, as in Fig. \ref{fig:capvel graphs}c. The most probable gravitational capture time for this event is either during aphelion or perihelion, consistent with \citet{2012Icar..218..262G} and \citet{2017Icar..285...83F}.

\paragraph{Comparison to Clark et al. (2016)}
In the study by \citet{2016DPS....4831106C}, they detected an 8.1-second fireball over the Czech portion of the European Fireball Network (EFN) with two high-resolution digital camera observatories. Given their observations, they determined that the detected event had a 92-98\% chance of being captured by the Earth before impact detection. The DFN event described here was about 5.3 seconds in duration and was detected by six high-resolution digital camera stations in South Australia (Fig. \ref{fig:map}). Despite a large amount of data collected of our event (6 cameras with $>$ 500 data points), the results varied significantly between model choices. Previous studies have demonstrated the sensitivity initial orbits can have to the choice of initial velocity method \citep{10.1093/mnras/sty1841}.  This is especially true for shallow events that penetrate deeper into the atmosphere where $v_{0}$ variations are more sensitive to model choice. The capture probabilities given for the EFN event are valid for the triangulation method that they used, but similar to our event, the use of a different triangulation method on their data may likewise find a reasonably high variation in the TC probability. Given that the event described in \citet{2016DPS....4831106C} was longer and shallower than the one described here, the $v_{0}$ variation due to model choice may cause more discrepancy in their $v_{0}$ estimates if fitting to the entire trajectory. Despite this, the \citet{2016DPS....4831106C} event has fewer observations, decreasing the sensitivity of model choice. This is because the $v_{0}$ distributions for multiple models have a higher chance of overlapping and possibly not causing as large of an issue with the discrepancy between models. 

If an object likely has a geocentric orbit, we further need to prove it is of natural origin and not from a human-made object. The event observed by the EFN recorded spectral data of the fireball and was able to conclude the object was conclusively natural. The event described here, on the other hand, may still have originated from an artificial source; however, this is very improbable given the pre-atmospheric orbit of the event.

In the future, the best way to confirm TC impact events would be by collecting more data prior to atmospheric entry using telescopes; which may come to fruition with the beginning of observations in 2022 by the LSST \citep{2008arXiv0805.2366I,2015IAUGA..2257052F}. In addition, if TCs can be detected far enough in advance, future sample return missions could target these objects as the delta-v for the mission could be extremely low relative to other asteroid sample return missions.

\section{Conclusions}
Based on our analysis, the event DN160822\_03 detected by the Desert Fireball Network has a high pre-impact capture probability, as large as $>95\%$ captured with our most statistically robust model. We find that the probable capture time, capture velocity, capture semi-major axis, capture NEO group, and capture mechanism all vary annually, with most captures occurring during Earth's aphelion or perihelion. This has been noted to some extent previously  \citep{2012Icar..218..262G,2017Icar..285...83F}, but most of the annual probability variations associated with the Earth's eccentricity found for this particular event have not been described before. We also discover that the probability of capture occurring as a result of a close lunar encounter varies according to the lunar month for this event. Although, this is probably due to the specific geometry of this event (i.e., low inclination, high eccentricity, geocentric apogee $\approx1LD$). Despite the large amount of data collected by our six cameras of the event, we can not say for certain what the pre-atmospheric orbit was due to the highly unpredictable nature of the system, and the chaotic scattering that occurs with every close encounter with the Moon and the Earth. We caution future analysis of possible TCO events to explore the effects of small variations in the initial conditions and various triangulation methodologies. Despite these uncertainties and chaotic elements, we can determine the probable origins of this event statistically to be 88.4\% Apollos, 6.2\% Amors, 2.9\% Atiras, and 2.5\% Atens. In a couple of years, more fireball events like this may be able to confirmed by additional telescopic observations like those from the LSST. 

\section{Acknowledgements}\label{sec:symbols} 
This work was funded by the Australian Research Council as part of the Australian Discovery Project scheme. 

This research made use of Astropy, a community-developed core Python package for Astronomy \citep{robitaille2013astropy}. Simulations in this paper made use of the REBOUND code which can be downloaded freely at http://github.com/hannorein/REBOUND.

\section{Summary of Definitions and Abbreviations} 
Within this study we followed the notation of \citet{2012Icar..218..262G} and \citet{2017Icar..285...83F} for consistency.
\begin{itemize}
\item SLLS - Straight Line Least Squares triangulation method with extended Kalman filter for velocity and error determination 
\item DTF - Dynamic Trajectory Fit triangulation and dynamic modelling method
\item TC - Temporarily-Captured. The sum of the total TCOs and TCFs
\item TCF - Temporarily-Captured Flyby. TC that has not orbited the Earth once
\item TCO - Temporarily-Captured Orbiter. TC that has orbited the Earth at least once 
\item Sputnik - Particle in integration that originates from the Earth
\item NES - Natural Earth Satellite
\item NEO - Near Earth Object 
\item UB - Unbound (i.e., not gravitationally captured by the Earth) 
\item 1H - Came within 1 lunar Hill Spheres of the Moon 
\item 3H - Came within 3 lunar Hill Spheres of the Moon
\item SC - Particles that are still captured by the end of the integration
\item LD - Distance from Earth to the Moon
\end{itemize}

\bibliography{tco}

\end{document}